# High-gain optical parametric amplification with a continuous-wave pump using a domain-engineered thin-film lithium niobate waveguide

MENGWEN CHEN,[1] CHENYU WANG,[1] KUNPENG JIA,[1,*] XIAO-HUI TIAN,[1,7] JIE TANG,[2] CHUNXI ZHU,[1] XIAOWEN GU,[2] ZEXING ZHAO,[1] ZIKANG WANG,[1] ZHILIN YE,[1,3] JI TANG,[1,3] YONG ZHANG,[1] ZHONG YAN,[3,4] XUEWEN WANG,[5] GUANG QIAN,[2] BIAOBING JIN,[1,6] ZHENLIN WANG,[1] SHI-NING ZHU,[1] AND ZHENDA XIE,[1,8]

[1]*National Laboratory of Solid State Microstructures, School of Electronic Science and Engineering, College of Engineering and Applied Sciences, School of Physics, Research Institute of Superconductor Electronics (RISE) & Key Laboratory of Optoelectronic Devices and Systems with Extreme Performances of MOE, Key Laboratory of Intelligent Optical Sensing and Manipulation, Ministry of Education, and Collaborative Innovation Center of Advanced Microstructures, Nanjing University, Nanjing 210093, China*
[2]*National Key Laboratory of Solid-State Microwave Devices and Circuits, Nanjing Electronic Devices Institute, Nanjing 210016, China*
[3]*Nanzhi Institute of Advanced Optoelectronic Integration, Nanjing 211800, China*
[4]*School of Integrated Circuits, Nanjing University of Information Science and Technology, Nanjing 210044, China*
[5]*State Key Laboratory of Advanced Technology for Materials Synthesis and Processing, Wuhan University of Technology, Wuhan 430070, China*
[6]*Purple Mountain Laboratories, Nanjing 211111, China*
[7]*tianxiaohui@nju.edu.cn*
[8]*xiezhenda@nju.edu.cn*
*\*jiakunpeng@nju.edu.cn*

**Abstract:** While thin film lithium niobate (TFLN) is known for efficient signal generation, on chip signal amplification remains challenging from fully integrated optical communication circuits. Here we demonstrate the continuous-wave-pump optical parametric amplification (OPA) using an x-cut domain-engineered TFLN waveguide, with high gain over the telecom band up to 13.9 dB, and test it for high signal-to-noise ratio signal amplification using a commercial optical communication module pair. Fabricated in wafer scale using common process as devices including modulators, this OPA device marks an important step in TFLN photonic integration.

## 1. Introduction

The recent development of information technology not only requires high-speed communication, but also calls for integrated optical communication devices, for applications including data center communication [1–3], satellite networks [4,5], integrated photonic processors [6,7], and optical integrated sensing and communication (O-ISAC) [8–10]. In recent years, thin film lithium niobate (TFLN) devices keep setting new records on the modulation speed and the low power consumption per bit [11–13], in a chip-scale footprint, which is beneficial for the further integration towards a compact optical communication device. Such integration not only relies on linear optical devices, as demonstrated using TFLN before [14–18], but also relies on the on-chip optical signal amplification. Because this optical signal amplification is to fulfill the fundamental signal-to-noise ratio (SNR) requirement against optical losses [19,20], following the general Shannon-Hartley theorem. The most popular off-chip solution for optical signal amplification is the use of erbium-doped fiber amplifiers (EDFAs) [21,22], which inevitably increases the size of the system. The similar concept can be adapted to on-chip devices, by the laser-active doping in TFLN. However, these laser-active devices are fabricated from doped lithium niobate crystals before the smart cut process so far [23–28], and thus the whole TFLN chip has to be doped with the same ion density, which limits the performances of non-laser-active devices including high-speed electro-optic modulators (EOMs).

On the other hand, optical parametric amplification (OPA) is an attractive alternate for the signal amplification [29,30], and has been known for its potential high gain and low noise for decades. The TFLN device confines light in a small mode field size for much stronger nonlinearity compared to bulk material [31,32], and thus offers the new opportunity for a high-gain OPA with significantly low peak power, i.e., using the continuous-wave (CW) pump. Such uninterrupted amplification capability lays the fundamental basis for practical modulated signal amplification. However, to date, only pulse-pumped OPAs have been demonstrated on TFLN [33,34], and the ability to achieve the TFLN-based OPA with a CW pump is still challenging [35–37]. The challenges mainly lie in the immature fabrication of domain-engineered TFLN waveguides, preventing the full exploitation of potential nonlinear optical performances.

Here we demonstrate the CW-pumped OPA on TFLN. Cascaded second-harmonic generation (SHG) and OPA processes are used for an indirect-pump geometry, which greatly simplifies pumping condition and coupling design. High on-chip parametric gain is measured with a broad 110 nm 10-dB bandwidth, covering both C and L bands at telecom wavelength, with gain up to 13.9 dB. Such OPA device has been tested for the communication signal amplification using 1.25 Gbps and 3.125 Gbps commercial communication modules, and results show successful signal amplification via OPA for lower bit error rate (BER) in comparison to the case without OPA. These results are achieved because of the high-quality fabrication of domain-engineered TFLN waveguides with high nonlinearity over long interaction length of 12.3 mm. The ion-beam trimming is the key to retain such high nonlinearity by suppressing the thickness variation and its side effects on the phase matching. Both waveguide fabrication and domain engineering are performed using deep-ultraviolet (DUV) lithography method over TFLN wafer, and share common processes as for high-performance EOMs and other wafer-scale devices for mass production. Therefore, these results add the on-chip amplification capability of TFLN photonics and mark an important step towards fully integrated TFLN optical communication circuits.

## 2. Device design and fabrication

We fabricate the OPA devices on an x-cut TFLN wafer, which is compatible with the fabrication of EOMs and other high-performance TFLN devices [13,38,39]. Instead of direct pumping, we use cascaded processes, where the pump is indirectly generated on chip via SHG process, so that only telecom-band fundamental wave (FW) input is necessary besides the signal light, as illustrated in Fig. 1(a). Equation (1) illustrates the cascaded SHG and OPA processes in our device:

$$2\omega_{FW,\,TE00} \rightarrow \omega_{p,\,TE00} \; (SHG),$$
$$\omega_{p,\,TE00} + \omega_{s,\,TE00} \rightarrow 2\omega_{s,\,TE00} + \omega_{i,\,TE00} \; (OPA), \qquad (1)$$

where $\omega_{j,TE00}$ ($j = FW, p, s, i$) corresponds to the angular frequency for FW, pump, signal and idler light, respectively. These two processes are designed to be phase-matched for the fundamental waveguide modes, i.e., $TE_{00}$ in this case. Both SHG and OPA can be quasi-phase matched (QPM) using the same period $\Lambda$ via domain engineering. Fig. 1(b) schematically illustrates the evolution of FW, SHG, signal and idler intensity along the waveguide for this cascaded process, showing SHG generation and then OPA process while the pump power increases from the SHG. With this scheme, the input coupling only need to be optimized for telecom band wavelength. More importantly, such indirect pumping scheme enables deterministic excitation for the short-wavelength pump light in fundamental modes as shown in Eq. (1). In a normal direct pumping scheme, however, it can be challenging for the coupling design to achieve high efficiency excitation of fundamental modes, considering the multi-mode feature of TFLN waveguide at short-wavelength pump.

It is also interesting to engineer a broad gain bandwidth of OPA, to fulfill the general requirement of the signal amplification. The gain bandwidth is determined by the phase matching [40,41]. For an interaction length of $l$, the phase mismatch $\Delta\varphi$ related to the frequency detuning $\Delta\omega$ can be given by

$$\Delta\varphi = l\left(k_p - k_s - k_i - \frac{2\pi}{\Lambda}\right) = -l\beta_2(\Delta\omega)^2 - 2l\sum_{m=2}^{\infty}\frac{\beta_{2m}}{(2m)!}(\Delta\omega)^{2m}, \qquad (2)$$

where $k_j$ ($j = p, s, i$) is the wave vector for pump, signal and idler light, respectively. In the first order approximation, it is inversely proportional to the group velocity dispersion (GVD) $\beta_2$. Thus, a lower GVD results in a larger gain bandwidth. In selecting the waveguide width, we consider the GVD as one of factors, due to its sensitive dependence on width.

The dispersion and QPM condition are sensitive to TFLN waveguide dimensions in cross-section. Specifically, the current TFLN wafers are normally fabricated with total thickness variation (TTV) on the order of tens-of-nanometers, and thus dramatically change the phase matching and reduce the parametric gain [42]. This makes it difficult to achieve high overall nonlinear efficiency ($P_{SHG}/P_{FW}^2$, $P_{SHG}$: SHG power, $P_{FW}$: FW power) from domain-engineered TFLN waveguides, even though high length-normalized nonlinear efficiency ($P_{SHG}/(P_{FW}^2 \times l^2)$) can be easily calculated using very short waveguides [43–46]. Recently, adaptive poling is presented as an effective method for high-efficiency SHG performances [47,48], by patterning variable poling periods according to the mapped geometric variation. In this work, we offer an alternative method, i.e., the ion-beam trimming [49–51], to actively improve the uniformity of film thickness over the whole wafer at nanometer level, so that high overall conversion efficiency can be achieved with simple periodic poling, and in a way that is fully compatible with wafer-scale fabrication process. A focused ion beam scans the TFLN wafer while sputter etching the material. Through coordinated control of the focused beam spot and dwell time at localized regions, spatially selective material removal is achieved to adjust film thickness, thereby effectively reducing film thickness variations. Usually, the thickness variation $\Delta t$ of the x-cut TFLN wafer is around 20 nm. By implementing the ion-beam trimming, the central area of wafer can be trimmed within sub-10 nm $\Delta t$, while the $\Delta t$ in edges can be too large from a perfect trimming, as shown in Fig. 1(d). Still, we achieve about 2 nm root mean square (RMS) value for $\Delta t$ along centimeter length in central area.

Efficient OPA with CW pump also requires both high nonlinearity and low optical loss. Normally, domain engineering is achieved in a poling-prior-etching fabrication process. However, this fabrication process normally suffers from selective etching problem during waveguide fabrication [52–54], where periodic steps can be formed along waveguide sidewalls, as a result of different etching rate following the existing domain structures. These periodic steps inevitably introduce extra poling-dependent scattering loss, and it can dominate when a long domain-engineered TFLN waveguide is used. We adopt a different approach, the etching-prior-poling process [48,55] to eliminate the selective etching problem. The fabrication procedure is shown in Fig. 1(c). The patterning of both waveguides and poling electrodes is achieved using DUV lithography in step-by-step process with high precision overlay alignment in-between, over an ion-trimmed, 4-inch, 600-nm-thick x-cut TFLN wafer.

The first step is for the TFLN waveguide fabrication, with the edge coupler on each end of the waveguide for efficient mode coupling. The edge coupler consists of a double-layer-etched, inversely-tapered TFLN waveguide, covered by a thick $SiO_2$ waveguide. Following the DUV patterning, argon ions reactive ion etching is performed to form the TFLN waveguide and the double-layer TFLN taper. The DUV exposure field size is 15 × 16 mm² for one shot, which results in 21 chips from a single wafer. The second step is for the domain engineering over existing waveguides. Advancing the strategy of our prior wafer-scale poling on thin film [56], we engineer its poling adaptation for waveguide structures. The dual comb-shaped poling electrodes along the ±z axis are designed, positioned away from waveguide sidewalls, and featuring a ~30% width duty cycle. The patterned electrodes are deposited with a 40-nm-thick Au layer stacked over a 30-nm-thick Cr layer, via electron-beam evaporation, and shaped in a lift-off process. The 1-ms-long, high-voltage pulses are applied with a poling electric field of ~100 kV/mm, significantly exceeding the coercive field to achieve domain inversion across the waveguide region. A photograph of wafer after electrode fabrication is shown in Fig. 1(e). Domain-engineered waveguides with high-quality poling and smooth sidewalls over domains are achieved as shown in Figs. 1(f) and 1(g). The duty cycle of domains inside the waveguide area is near 50:50, with a small period of 4.17 μm. After removing the poling electrodes, a 2-μm-thick $SiO_2$ layer is then deposited via plasma-enhanced chemical vapor deposition, serving as the upper cladding. Following UV patterning, it is etched to form the $SiO_2$ waveguide in the edge coupler region. Chips are finally separated by dicing, grinding, and polishing based on alignment marks. We measure the dimensions in the waveguide cross-section for thickness $t$, top width $w$ and slab thickness $s$ of 594 nm, 1210 nm and 311 nm, respectively, and $TE_{00}$ mode profiles are simulated using measured dimensions, as shown in Fig. 1(h). The dispersion $β_2$ is then calculated to be ~0.28 $ps^2/m$ (@ 1550 nm) around the signal wavelength, as shown in Fig. 1(i).

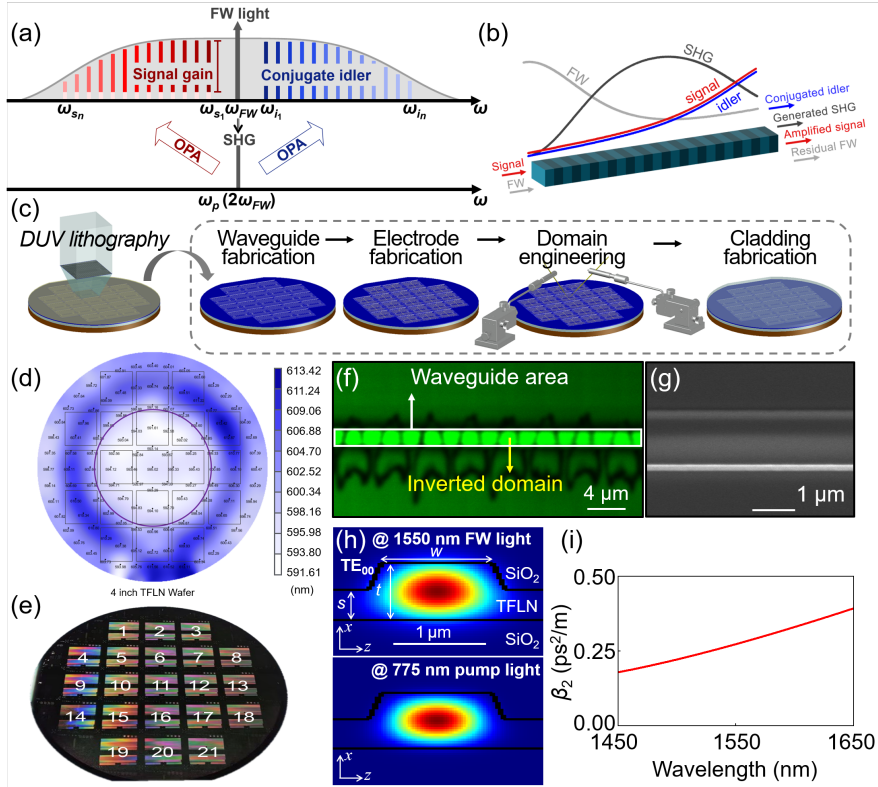

Fig. 1. (a) Diagrammatic depiction of the cascaded SHG and OPA processes. The on-chip pump is indirectly generated through the SHG process. Besides the signal light, only the telecom-band FW input is required. (b) Schematic of FW, SHG, signal, and idler intensities evolution along the waveguide during the cascaded process. (c) Wafer-scale fabrication procedure of domain-engineered TFLN waveguides. The waveguides are fabricated prior to the domain engineering process. (d) Thickness map of the ion-beam trimmed TFLN wafer. The exposure positions of chips across the wafer, and the central area of TFLN with lower thickness variation are marked. (e) Photograph of the wafer after electrode fabrication. The 21 chip numbers are labeled on the photograph. (f) Confocal microscope image for domain-engineered waveguide. The waveguide area is highlighted by the white frame, and the inverted domain is indicated by the yellow arrow.

(g) SEM image of the waveguide from the top view. (h) Schematic of the waveguide cross section in x-cut TFLN, with the simulated $TE_{00}$ mode profiles for FW light and pump light. (i) Calculated waveguide dispersion $\beta_2$ around the signal wavelength.

## 3. Results

We first measure the insertion losses of waveguides in a fiber-in and fiber-out setup, for ones with and without domain-engineering in comparison. Both types of waveguides show consistent insertion losses of 4.1 ± 0.3 dB in C-band. Such insertion losses stay relatively stable for waveguides with different poling section lengths varying from 0 mm to 12.3 mm. The 12.3-mm-long domain-engineered waveguides are designed for the OPA experiment. Using a low-power CW tunable laser (TSL-550, Santec) as the FW light, we measure their SHG performances in comparison from chip to chip. Fig. 2(a) shows typical SHG measurement results from waveguides at different locations of the same wafer. In chips #6, #10, and #11 with low $\Delta t$ from the central part, peak normalized efficiencies are measured from 1610 to 1790 %/W/cm², with corresponding peak overall efficiencies from 2440 to 2710 %/W. While in chips #1 and #3 with much higher $\Delta t$ from the edge part, much lower efficiencies are measured at about 1600 %/W. We choose the waveguide from chip #10 for the OPA experiment. By measuring the quality factor ($Q$) of the micro-ring resonators with the same waveguide structure of the OPA device, the propagation loss is estimated to be on the order of 0.1 dB/cm. For the OPA waveguide, the SHG power measurement while changing the FW power is shown in Fig. 2(b). The normalized SHG efficiency of 2670 %/W is calculated from a quadratic fitting marked in the blue curve.

Strong on-chip pump power is required for high-gain OPA. We use an EDFA to boost the power of FW light and measure its depletion ratio at the output side, for the indirect measurement of on-chip SHG efficiency. The depletion ratio is measured by quantifying relative power of the residual FW light at phase-matching wavelength versus phase-mismatching wavelengths, achieved by rapidly scanning the FW wavelength, so that the measured ratio is not affected by the coupling loss. The relative powers are measured using the photodetector. With the off-chip FW power set at 33.6 dBm, its depletion ratio as a function of FW wavelength is shown in Fig. 2(c). A maximum depletion of about 90% can be measured at the phase matching wavelength of 1572 nm, indicating a maximum on-chip SHG efficiency of about 90%.

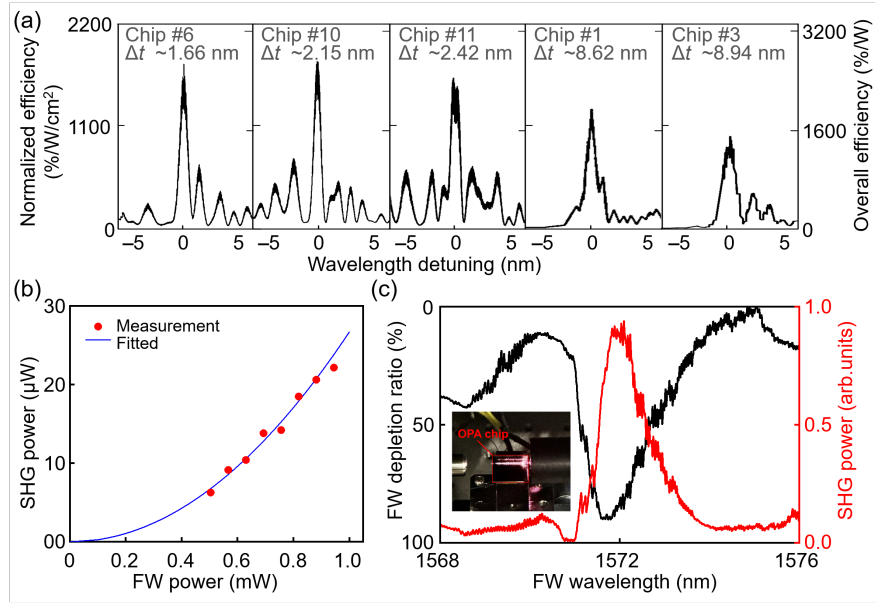

Fig. 2. (a) Typical SHG results of 12.3-mm-long waveguides from different chips, with different thickness variation $\Delta t$. The waveguide from chip #10 is chosen for the OPA demonstration. (b) SHG power as a function of input FW power measured from the OPA waveguide. (c) FW depletion ratio and generated SHG power as a function of FW wavelength, indicating strong on-chip SHG generation. The inset shows the picture of the OPA chip with the shining waveguide. The OPA chip is marked by a red frame.

Then we are ready for the OPA experiment. The setup is schematically shown in Fig. 3(a). The OPA chip is placed on a temperature-controlled metal mount, with temperature stability of 2 mK. The FW light is from the amplified tunable laser as discussed above, and a pair of narrow bandpass filters are used to filter out the amplified spontaneous emission noise from the EDFA. The FW light is then combined with the signal light on a dichroic mirror, and coupled into the OPA chip with an aspherical lens. Two sets of $\lambda/2$ and $\lambda/4$ waveplates are used to independently adjust the polarization of FW and signal light, respectively. At the output side, we use a lensed fiber to collect the output of OPA chip. For the gain measurement, the narrowband bandpass filter at signal wavelength is used before the power meter. For the spectral analysis, the lensed fiber is connected to a directional coupler (99:1), with its 1% output port to a silicon photodiode for monitoring the residual pump light, and 99% output port to an optical spectrum analyzer (OSA, AQ6374, YOKOGAWA). The spatial grating filter system with isolation over 80 dB, is inserted before the OSA to prevent any residual FW light from affecting the spectral detection.

The OPA performance is first tested with a CW signal input from a tunable laser (TSL-550, Santec). With only FW light input, the power at signal wavelength with narrow bandpass filtering is lower than the power meter background noise. We turn off the FW light and couple only the signal light into the OPA chip. The output spectrum of OPA chip is recorded using OSA, as shown in Fig. 3(b) (solid brown curve). We then turn on the FW light with an off-chip power of 33.6 dBm. Obvious increase in signal power is observed in the spectrum, indicating the occurrence of amplification (Fig. 3(b)). By comparing the signal power with and without the FW light, i.e., with and without OPA, the on-chip gain can be derived to be 13.9 dB, corresponding to a net gain of 9.9 dB with the 4 dB insertion loss of the OPA device. The on-chip gain normalized by device length and FW power features 0.78 dB/(W·mm). The generated conjugate idler light is also observed in spectrum. The power level of the idler light is measured to be comparable to that of the signal light (< 0.2 dB), due to the intense OPA. We further measure the optical signal-to-noise ratio (OSNR) of the amplified signal light under different bandwidth settings, to simulate channels with different baud rates. Within the bandwidth ranging from 20 GHz to 200 GHz, the values of OSNR consistently exceed 28 dB. For the 20 GHz bandwidth setting, the OSNR is ~36 dB, corresponding to a noise figure [57,58] of ~2.1 dB. The lower noise of our OPA device, compared to traditional amplifiers like EDFAs [59,60] or SOAs [61,62], is expected to achieve higher-fidelity amplification in communication systems, for better performance metrics such as the bit error rate (BER) [20,63].

By fixing the wavelengths of FW light and signal light, we measure the gain at different FW light power, as shown in Fig. 3(c). The OPA gain also varies with the signal wavelength due to the change of the phase matching condition. We measure the on-chip gain at different signal wavelength in the range from 1520 nm to 1630 nm. The results are shown in the upper panel of Fig. 3(d), with the corresponding spectra shown below. The measured 10-dB bandwidth of OPA is over 100 nm, covering the whole telecom C and L bands, which is attributed to our dispersion engineering. The linewidth of the CW-pumped FW light only occupies a small portion in the large gain bandwidth, forming a narrow rejection band, which is of importance for wideband signal amplification. In this high-gain OPA process, vacuum quantum fluctuations are amplified to a macroscopic level through optical parametric generation (OPG). This results in a flat spectrum of OPG at the base of spectra in Figs. 3(b) and 3(d).

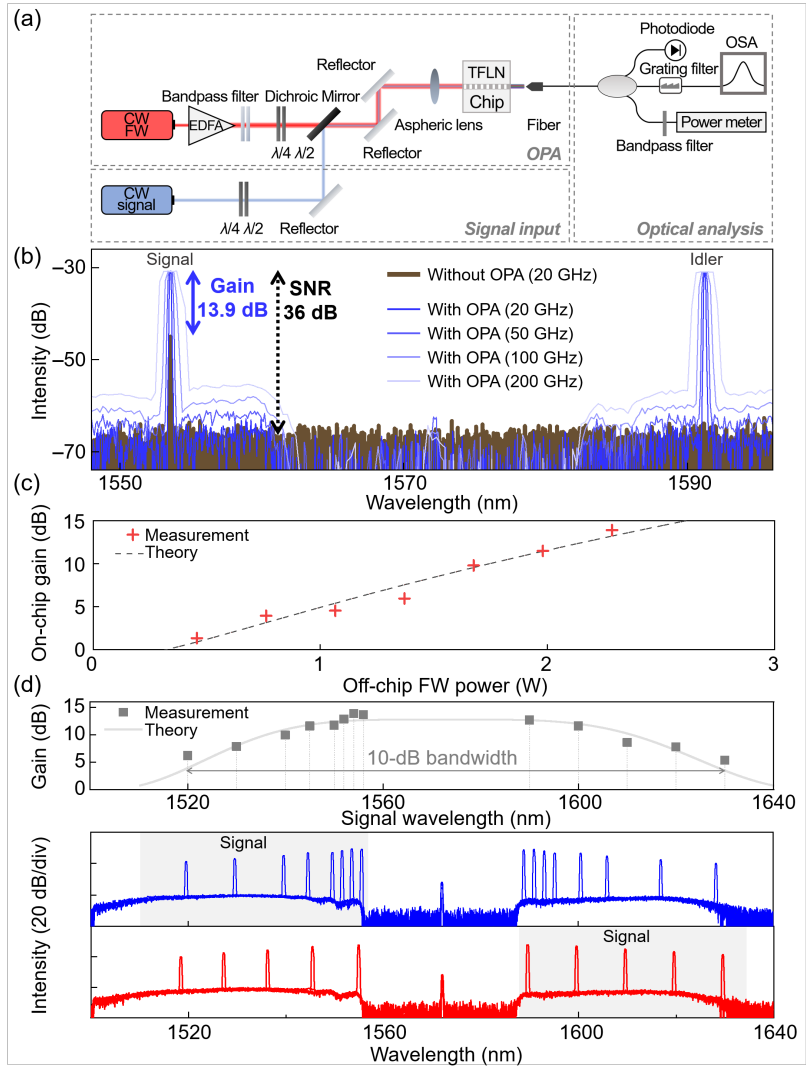

Fig. 3. (a) Experimental setup for CW-pumped OPA. (b) Output spectra of OPA chip, with only signal light on (without OPA), and with signal and FW lights on (with OPA) in different optical bandwidths. (c) Measured on-chip gain as a function of off-chip FW light power. (d) On-chip gain as a function of signal wavelength, with theoretical (solid grey line) and measured (grey dots) data. The corresponding spectra are shown below. The spectra on the signal light side are marked with a grey background.

It is interesting to test the amplification of the modulated signal using our OPA device, serving as a proof of concept for future applications. The modulated signals are from commercial communication modules with two different data rates at 1.25 Gbps and 3.125 Gbps. The experimental setup is illustrated in Fig. 4(a). The modulated signal light, from Alice (the transmitter), goes through our OPA device and single mode fiber (SMF) to Bob (the receiver). A high-speed oscilloscope (MSO 71604C, Tektronix) is used for eye diagram analysis, thus the BER values can be derived. The presence or absence of OPA gain is controlled by the off-chip FW input. When the FW light is injected to the OPA chip, the modulated signal light can be amplified. We also perform a controlled experiment using low-noise EDFA for signal amplification instead of OPA, for comparison.

The BER performance is tested with different input signal power, as illustrated in Fig. 4(b), for cases with OPA, with EDFA and without amplification. Notably, with OPA gain, the signal power as low as –44 dBm are elevated to be above the detection threshold for sensitivity improvement. In contrast, without amplification, the sensitivity of optical modules is limited at –32 dBm, restricting communication capabilities for signals with lower power. The amplified signals with OPA gain consistently show lower BER results, than that of the cases without amplification, especially at the power level over the detection threshold. The BER decreases to the low level of $10^{-14}$, with the increase of the amplified signal power to the –14 dBm level. In comparison to that of the cases with EDFA, the performance of the case using OPA is better at the same power level. These results underscore the exceptional low-distortion amplification capabilities of our OPA device, attributed to its high gain and low noise. Note that the electro-optic modulation for data encoding can be further realized by on-chip EOMs, and integrated with our OPA device towards fully integrated communication circuits.

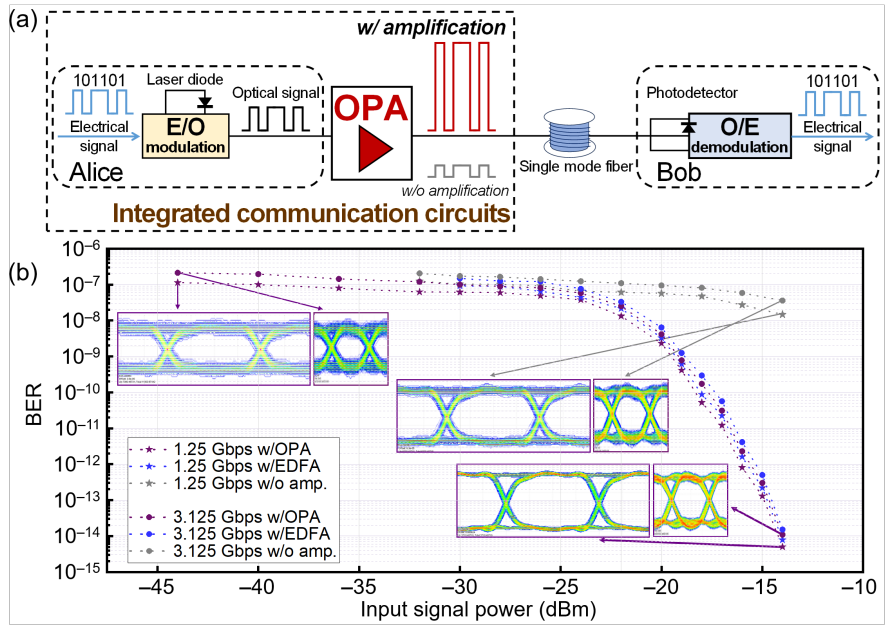

Fig. 4. (a) Experimental setup for modulated signal amplification. A pair of commercial optical communication modules serve as Alice (the transmitter) and Bob (the receiver), respectively. The OPA device can be further integrated with on-chip EOMs towards fully integrated communication circuits. (b) BER at different input signal power, for cases with OPA, with EDFA and without amplification.

## 4. Conclusion

In summary, we report the high-gain OPA with CW pump using the TFLN waveguide. With up to 13.9 dB on-chip gain over a broad telecom band, high-fidelity amplification for modulated signal is performed. These results mark a cornerstone for fully integrated optical communication circuits. The OPA device is fabricated via our state-of-the-art wafer-scale fabrication of domain-engineered TFLN waveguides, which demonstrates a critical combination of high nonlinearity, low loss, broad bandwidth and high-power handling capability, being ready for further integration with other TFLN devices, such like high-performance EOMs [38,39,64].

Currently, our OPA device requires watt-level off-chip FW power. The major limit on reducing power consumption is from the existing few nanometer-level TTV over TFLN wafer, which prevents us from using longer waveguides for higher overall efficiency in this demonstration. However, this is not a fundamental limit because the better TTV can be expected by improving the smart-cut technology or further developing the higher-precision ion-beam trimming technique. In a 100-mm-long TFLN waveguide with $\Delta t$ of sub-1 nm, our simulation shows that power consumption below 100 mW is enough for even higher gain over 40 dB, which is within the power budget over single-mode semiconductor laser diode. A pump-integrated OPA device can thus be expected to utilize hybrid integration technology. Meanwhile, promising approaches including tunable Vernier filters [65,66], photonic crystal filters [67,68] and Bragg gratings [69,70] are also expected for future on-chip filtering. Also, we can further demonstrate phase sensitive amplification (PSA) [71,72] to fully reveal the high SNR advantage below the traditional quantum limit, which is not available using normal amplifiers like EDFAs [59,60] and SOAs [61,62].

The highly efficient, low-loss nonlinear device we have developed, either standalone or integrated, not only serves as an indispensable OPA device for numerous photonics applications in optical communication [73–76], photonic processors [77,78] and photonic computing [79–82], but also provides a pivotal solution for extensive fundamental research, including quantum optics [83,84], precision spectroscopy [85,86], biophotonics and imaging [71,87].

**Funding.** National Key Research and Development Program of China (2022YFA1205100, 2023YFB2805700); National Natural Science Foundation of China (62288101, 62293523, 62293520, 12304421, 12341403, 92463304, 92463308); Zhangjiang Laboratory (ZJSP21A001); Guangdong Major Project of Basic and Applied Basic Research (2020B0301030009); Program of Jiangsu Natural Science Foundation (BK20230770, BK20232033).

**Disclosures.** The authors declare that there are no conflicts of interest related to this article.

**Data availability.** Data underlying the results presented in this paper are not publicly available at this time but may be obtained from the authors upon reasonable request.